\documentclass[seceq]{ptptex}

\usepackage{graphicx}

\def\lsim{\raise0.3ex\hbox{$<$\kern-0.75em\raise-1.1ex\hbox{$\sim$}}}
\def\gsim{\raise0.3ex\hbox{$>$\kern-0.75em\raise-1.1ex\hbox{$\sim$}}}
\title{
QCD thermodynamics at zero and finite densities \\ with improved Wilson quarks%
}

\author{
K.\ \textsc{Kanaya}$^a,$\footnote{Speaker},  
S.\ \textsc{Aoki}$^a$, 
S.\ \textsc{Ejiri}$^b$, 
T.\ \textsc{Hatsuda}$^c$, 
N.\ \textsc{Ishii}$^c$, \\
Y.\ \textsc{Maezawa}$^d$, 
H.\ \textsc{Ohno}$^a$, 
H.\ \textsc{Saito}$^a$, 
N.\ \textsc{Ukita}$^e$, 
T.\ \textsc{Umeda}$^f$
\\
(WHOT-QCD Collaboration)
}

\inst{
$^{a}$Grad.\ School of Pure and Applied Sci., Univ.\ of Tsukuba, Tsukuba 305-8571, Japan
\\
$^{b}$Grad.\ School of Sci.\ and Tech., Niigata Univ., Niigata 950-2181, Japan 
\\
$^{c}$Dept.\ of Physics, The Univ.\ of Tokyo, Tokyo 113-0033, Japan
\\
$^{d}$En'yo Lab., RIKEN Nishina Accel.\ Research Center, Saitama 351-0198, Japan
\\
$^{e}$Center for Comput.\ Physics, Univ.\ of Tsukuba, Tsukuba 305-8577, Japan
\\
$^{f}$Grad.\ School of Education, Hiroshima Univ., Hiroshima 739-8524, Japan
}

\abst{
The WHOT-QCD Collaboration is pushing forward lattice studies of QCD at finite temperatures and densities using improved Wilson quarks. 
We first present results on QCD at zero and finite densities with two flavors of degenerate quarks ($N_F=2$ QCD) adopting the conventional fixed-$N_t$ approach. 
We then report on the status of a study of $N_F=2+1$ QCD adopting a fixed-scale approach armed with the $T$-integration method which we have developed. 
}

\begin{document}

\maketitle

\section{Introduction}
\label{sec:intro}

Clarification of thermodynamic properties of hot/dense quark matter is important in the studies of early Universe and relativistic heavy ion collisions. 
Because the issue is essentially non-perturbative, numerical studies on the lattice is so far the only systematic way to investigate it directly from the first principles of QCD.
Most lattice studies of hot/dense QCD have been done with computationally cheap staggered-type lattice quarks. 
However, their theoretical basis such as locality and universality are not well established. 
Therefore, to evaluate the effects of lattice artifacts, it is important to compare the results with those obtained using theoretically sound lattice quarks, such as the Wilson-type quarks. 

The WHOT-QCD Collaboration is pushing forward a series of lattice studies using clover-improved Wilson quarks coupled to RG-improved Iwasaki glues.
A reason that Wilson-type quarks have not been intensively studied in hot/density QCD is that the computational cost is larger than that for staggered-type quarks, in particular at small quark masses. 
Therefore, the previous studies with Wilson-type quarks were limited to the cases of two flavor QCD with heavy quarks at vanishing chemical potentials.
We want to extend the studies to more realistic $2+1$ flavor QCD at finite chemical potentials with physical light quarks.
Towards this goal, we made a series of simulations by implementing and developing efficient methods for Wilson-type quarks.

\section{Two flavors of improved Wilson quarks at $\mu=0$ and $\ne0$}
\label{sec:Nf2}

A systematic study of finite temperature QCD with clover-improved Wilson quarks was made by the CP-PACS Collaboration around the beginning of this century for the case of two flavor QCD at vanishing chemical potentials\cite{CPPACS00,CPPACS01}. 
It was noted that improvement is important for both gauge and quark actions to reproduce the O(4) scaling expected around the pseudocritical temperature near the chiral limit of two flavor QCD with unimproved Wilson quarks\cite{O4scaling}.
Therefore, they adopted the RG-improved Iwasaki gauge action\cite{Iwasaki} for gluons.
The study was made for quark masses in the range $m_{\rm PS}/m_{\rm V}$ \gsim\ 0.65 around the pseudocritical temperature on $N_t=4$ and 6 lattices.
They confirmed the O(4) scaling of subtracted chiral condensate and calculated the equation of state (EOS) along lines of constant physics (LCP).


The WHOT-QCD Collaboration is a successor of the CP-PACS Collaboration and has extended the studies to heavy quark screening masses\cite{WHOT07,WHOT10screening} and finite chemical potentials\cite{WHOT10dense}.
Because Wilson-type quarks are numerically more intensive, we had to adopt and develop several improvement techniques. 

   \begin{figure}[b]
       \centerline{
       \includegraphics[width=4.5 cm]{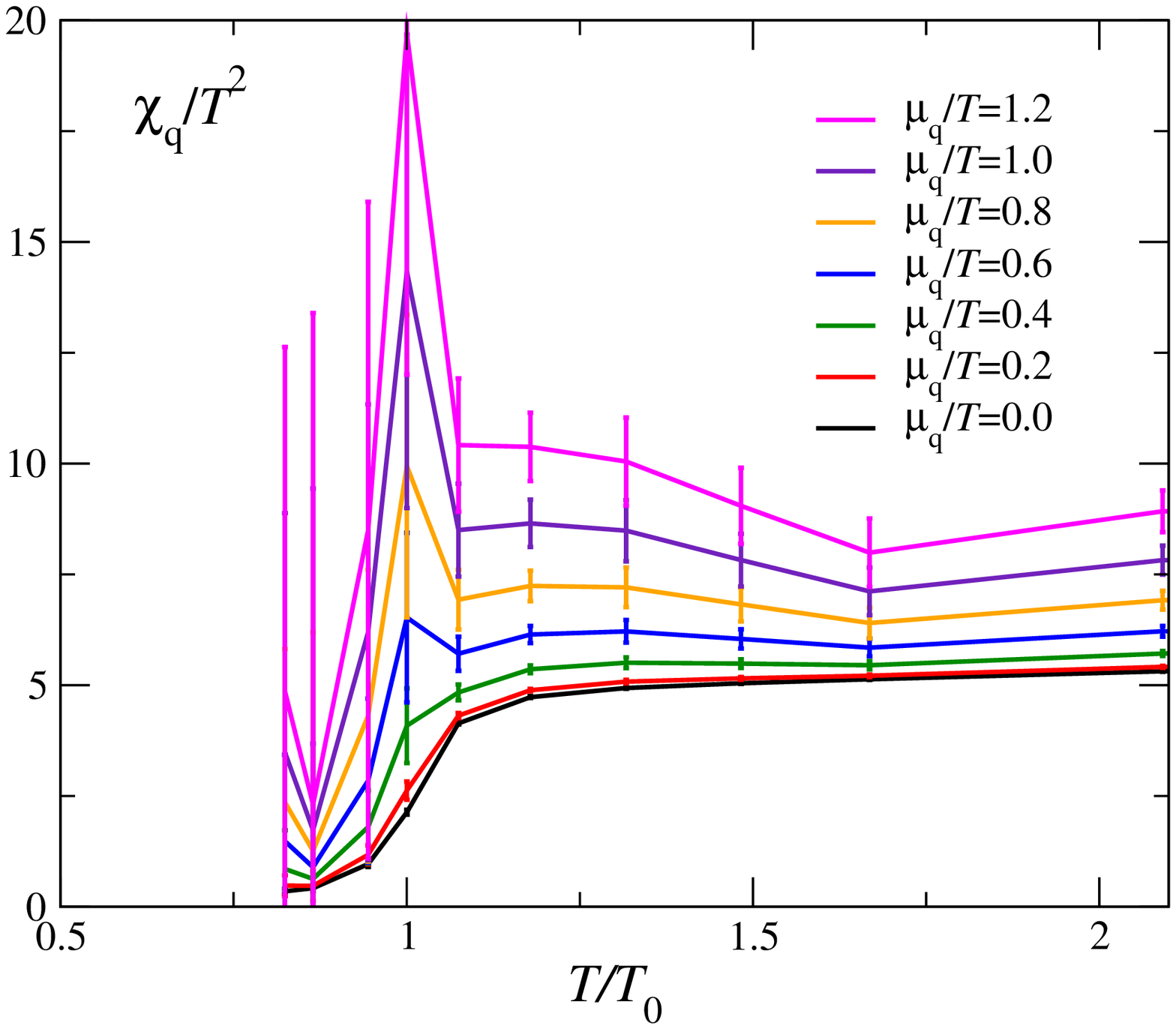}
       \hspace{1mm}
       \includegraphics[width=4.5 cm]{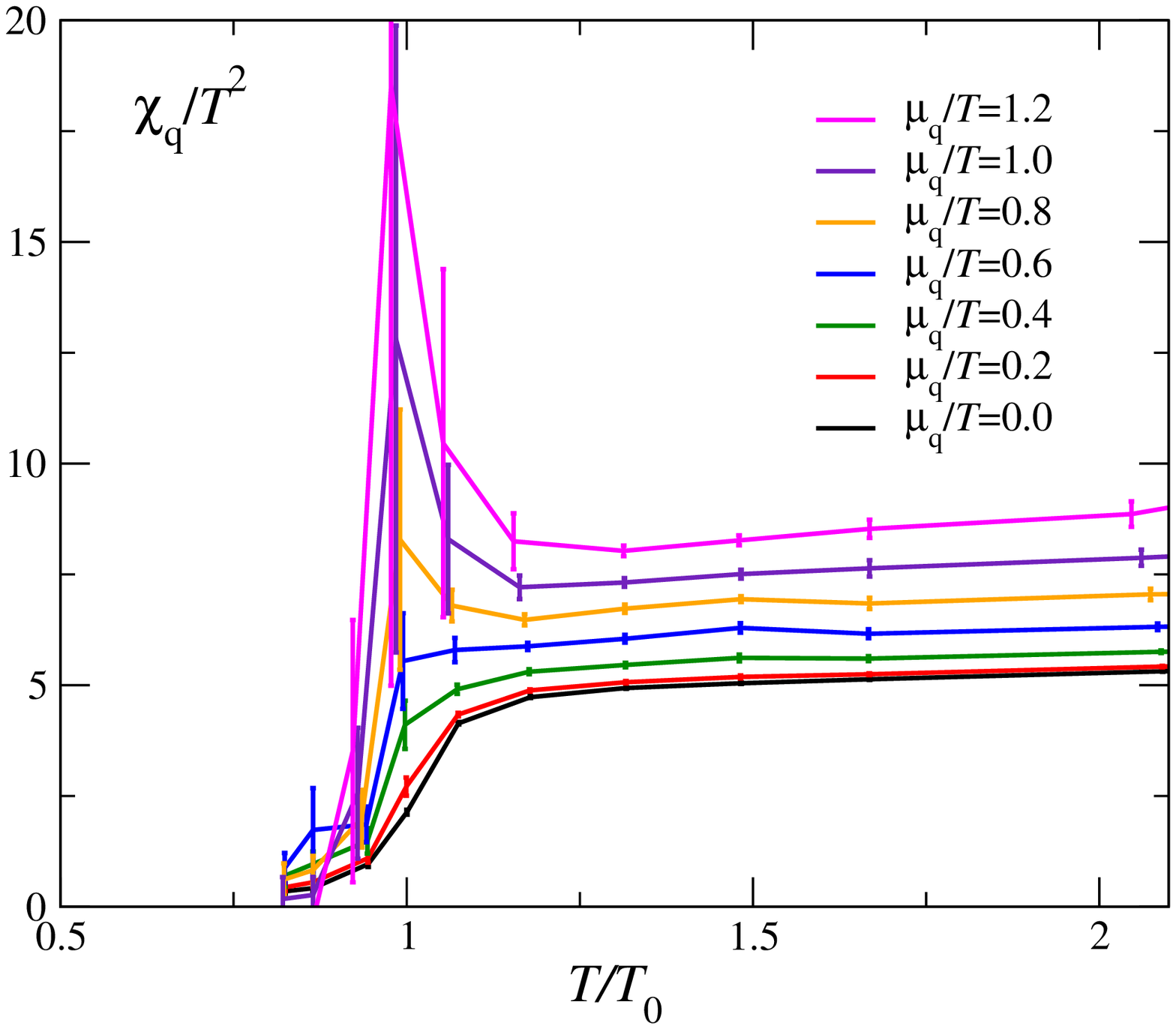}
       \hspace{1mm}
       \includegraphics[width=4.5 cm]{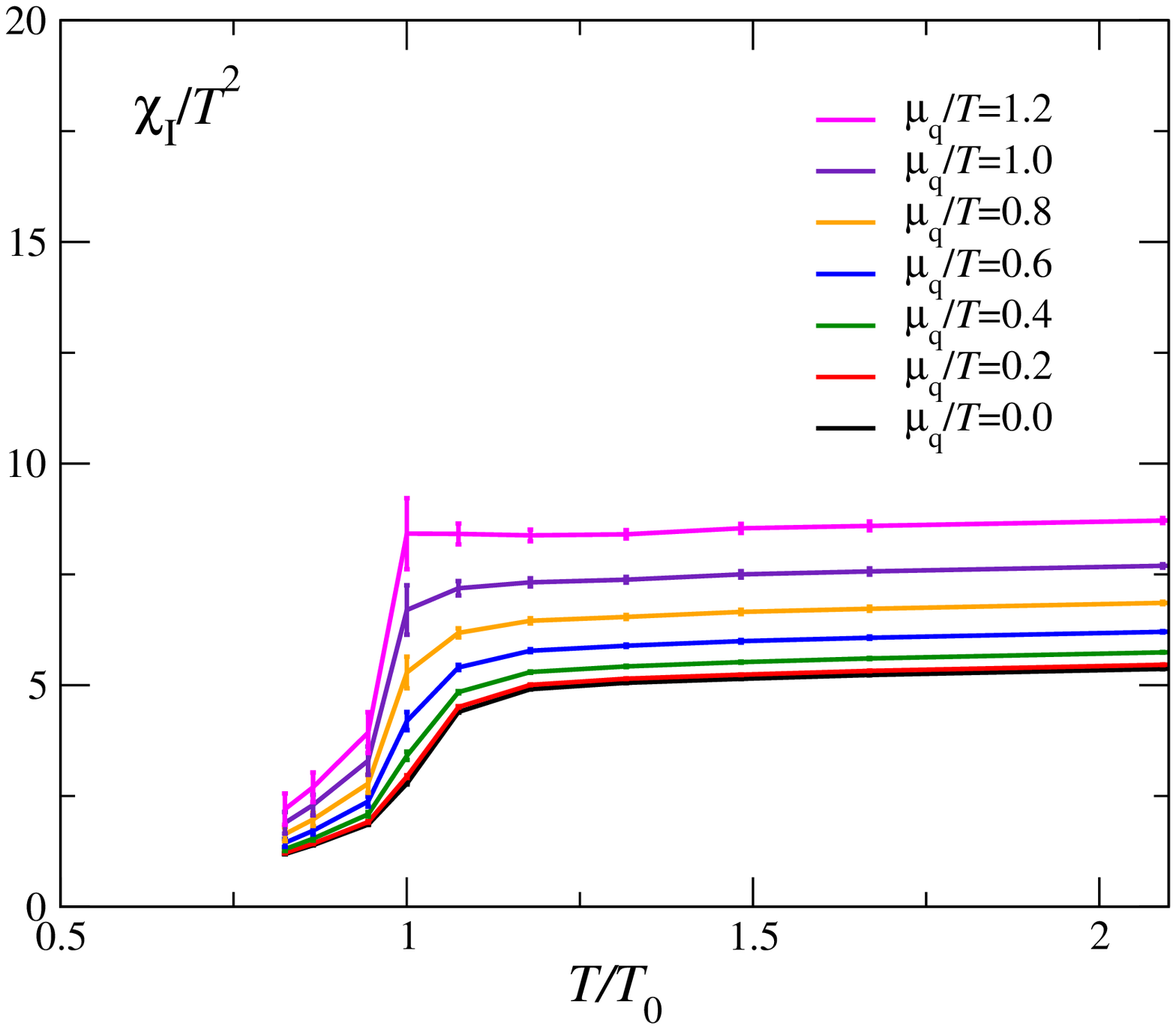}
       }
   \caption{Quark number susceptibility $\chi_q$ and isospin susceptibility $\chi_I$ for $m_{\rm PS}/m_{\rm V}=0.65$ at finite chemical potentials $\mu_q$\cite{WHOT10dense}. 
   {\em Left:} $\chi_q$ by the standard Taylor expansion method. 
   {\em Center:} improved results by the Gaussian approximation method. 
   {\em Right:} $\chi_I$ by the standard Taylor expansion method. 
   $T_0$ is the pseudocritical temperature at $\mu_q=0$.}
   \label{fig:wdense}
   \end{figure}

In Fig.\ref{fig:wdense}(left), we present the results of the quark number susceptibility $\chi_q(\mu_q)$ at $m_{\rm PS}/m_{\rm V}=0.65$ for small chemical potentials, obtained by using the standard Taylor expansion method up to the order $\mu_q^4$.
We find that the statistical errors are large in spite of various improvements in random noise estimators etc.
Therefore, we further apply a hybrid method of Taylor expansion and spectral reweighting, and introduce a Gaussian method to suppress the errors due to the complex phase fluctuation of the quark determinant\cite{Gaussian}.
The results shown in Fig.\ref{fig:wdense}(center) show suppression of statistical errors and smooth and natural $T$-dependence, although simulations at different temperature are independent.
We find that a sharp peak in $\chi_q/T^2$ appears near the pseudocritical temperature at finite $\mu_q$
and becomes higher as $\mu_q$ increases.
These are consistent with the observations with staggered-type quarks and suggest a critical point at finite $\mu_q$.
On the other hand, the isospin susceptibility shown in Fig.\ref{fig:wdense}(right) has no sharp peaks, 
in accordance with the expectation that $\chi_I$ is analytic at the critical point since the iso-triplet mesons remain massive. 
Results at $m_{\rm PS}/m_{\rm V}=0.80$ are similar, but the peaks in $\chi_q$ are much milder than those in Fig.\ref{fig:wdense}. 
This may be explained in part  by the expectation that the critical point locates at larger $\mu_q$ because the quark mass is larger than that for $m_{\rm PS}/m_{\rm V} = 0.65$.
See Ref.\citen{WHOT10dense} for more discussions.

\section{Fixed scale approach with the $T$-integration method}
\label{sec:FixedScale}

In order to extend the studies of the previous section to lighter quarks and to the case of $2+1$ flavor QCD, further reduction of the computational cost is required.
In conventional studies on the lattice, temperature $T=(N_t a)^{-1}$ is varied by changing the lattice scale $a$ through a variation of the lattice gauge coupling $\beta$, at a fixed temporal lattice size $N_t$.
In this fixed-$N_t$ approach, EOS is calculated by combining a calculation of the trace anomaly $\epsilon-3p$, where $\epsilon$ is the energy density and $p$ the pressure, and a non-perturbative estimates of $p$ using the integral method. 
Here, we note that a sizable fraction of the computational cost is devoted for $T=0$ simulations to set the basic parameters such as the lattice scale, to determine LCP's and the beta functions on them, and to carry out zero-temperature  subtractions for the renormalization of finite-temperature observables at each simulation point.

   \begin{figure}[b,t]
       \centerline{
    \includegraphics[width=46mm]{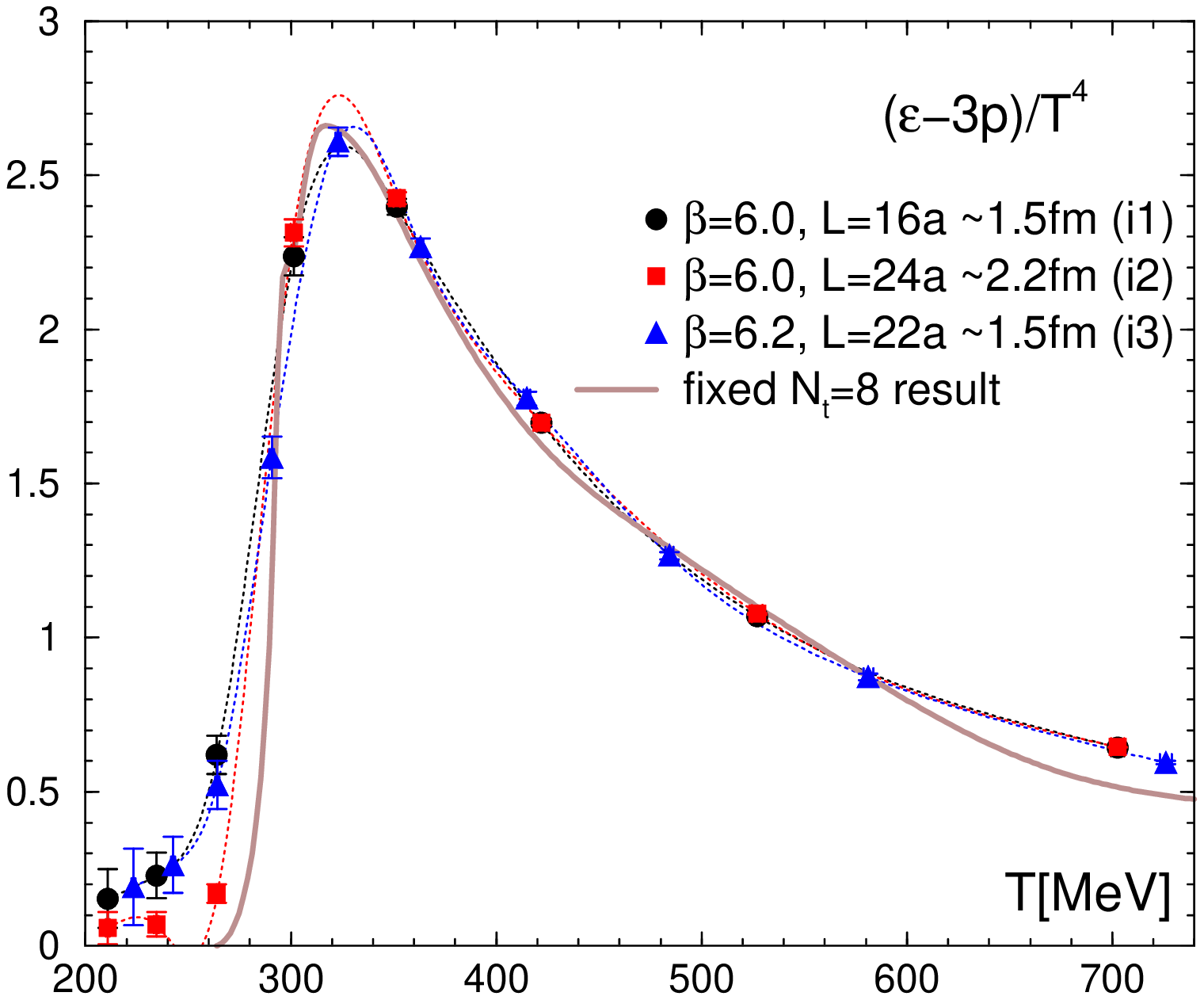}
    \hspace{0.5mm}
    \includegraphics[width=45mm]{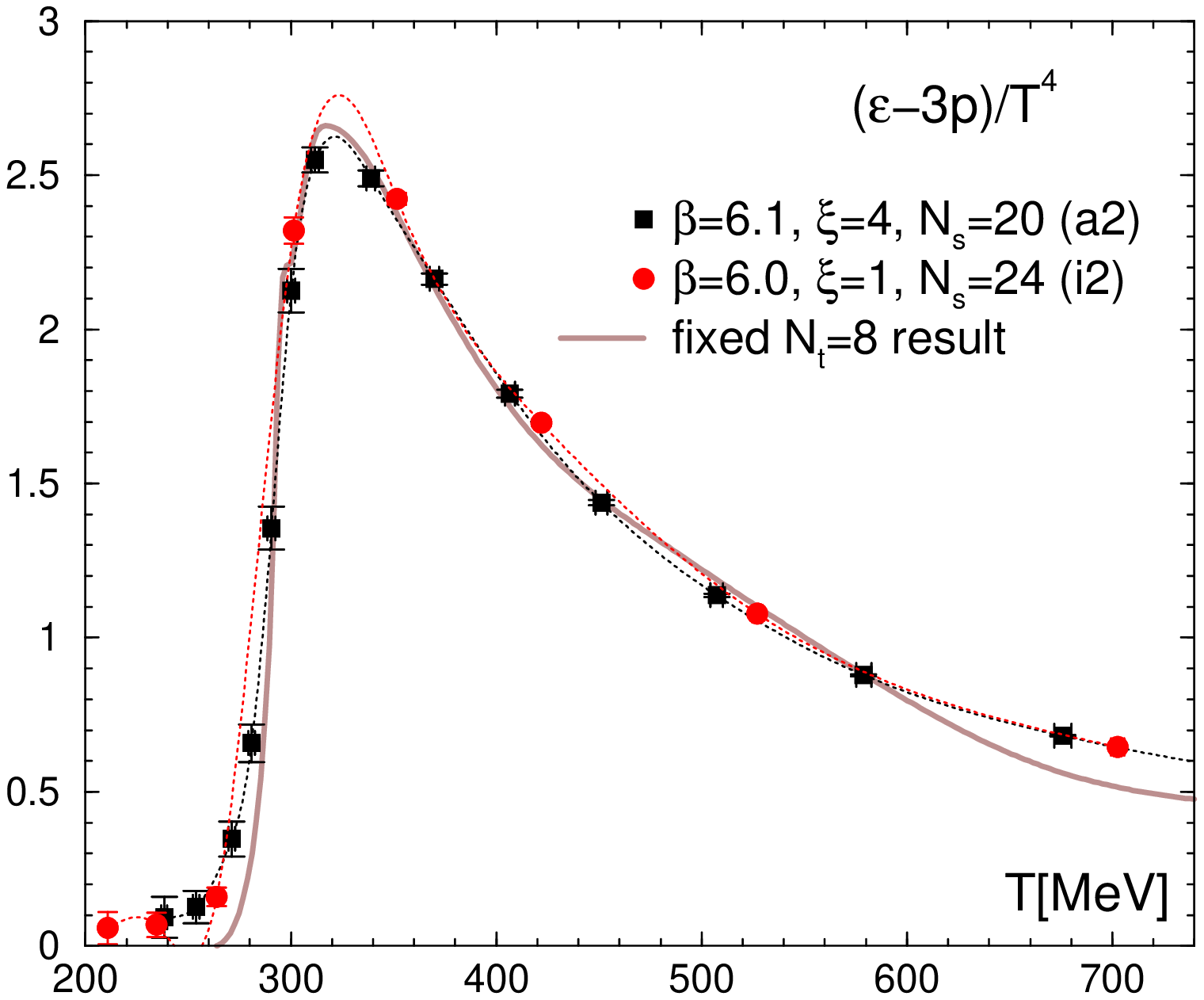}
    \hspace{1.5mm}
    \includegraphics[width=44mm]{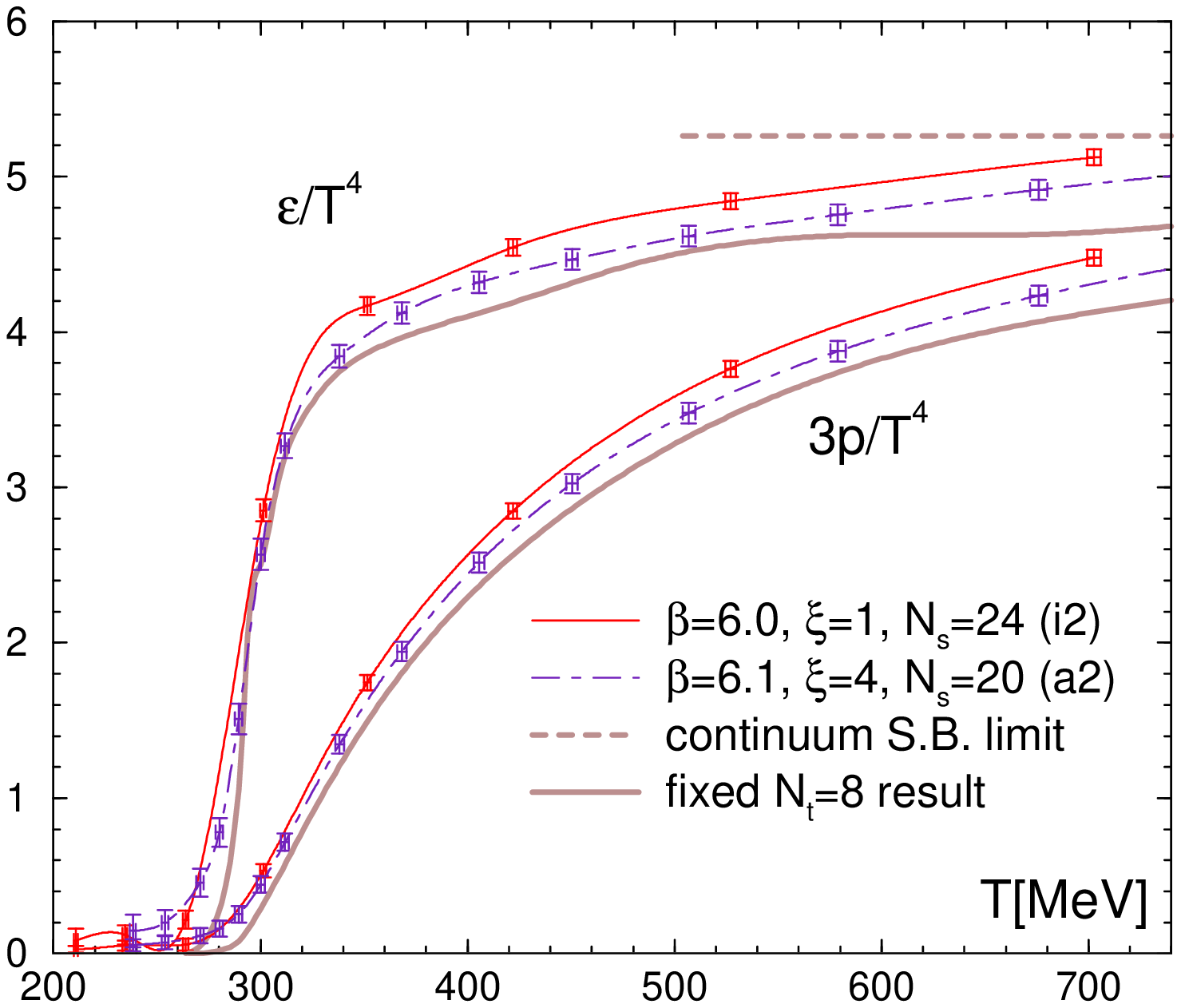}
       }
   \caption{Test of the fixed-scale approach armed with the $T$-integration method in quenched QCD\cite{WHOT09}. 
    {\em Left:} trace anomaly on isotropic lattices with different lattice spacing and volume.
    Dotted lines are the natural cubic spline interpolations of the data.
    {\em Center:} trace anomaly on an anisotropic lattice (a2) compared with the isotropic lattice with similar spatial lattice spacing and volume (i2).
    {\em Right:} energy density and pressure by the T-integral method.
    The shaded curves represent the results of the conventional fixed-$N_t$ method at $N_t= 8$ \cite{Boyd:1996bx}.}
   \label{fig:fixedscale}
   \end{figure}

To reduce the simulation cost, we proposed a {\em fixed-scale approach} to calculate EOS\cite{WHOT09}.
In this approach, the temperature $T$ is varied by $N_t$ with the lattice coupling parameters fixed (and thus with the lattice spacing $a$ fixed). 
Since the conventional integral method to obtain the pressure by an integration in the coupling parameter space is inapplicable, 
we developed a new method, {\em the $T$-integration method}, 
to calculate $p$ non-perturbatively:
Using a thermodynamic relation valid at vanishing chemical potential, we obtain
\begin{eqnarray}
T \frac{\partial}{\partial T} \left( \frac{p}{T^4} \right) =
\frac{\epsilon-3p}{T^4}
\hspace{6mm}\Longrightarrow\hspace{6mm}
\frac{p}{T^4} = \int^{T}_{T_0} dT \, \frac{\epsilon - 3p}{T^5}
\label{eq:Tintegral}
\end{eqnarray}
with $p(T_0) \approx 0$.
We thus evaluate the numerical integration in the right hand side.
Note that the resolution in $T$ is limited due to the discreteness of $N_t$.
Therefore, we need to check the magnitude of systematic errors from the interpolation of the trace anomaly in $T$.

Because (i) the $T=0$ subtractions can be done by a single $T=0$ simulation 
and (ii) all the simulations are automatically on a LCP, 
the fixed-scale approach enables us to largely reduce the cost for $T=0$ simulations.
We may even borrow high statistic configurations at $T=0$ on the International Lattice Data Grid.

We find that the fixed-scale approach is complemental to the conventional fixed-$N_t$ approach in many respects:
In the high $T$ region, typically $T \gg T_c$ where $T_c$ is the (pseudo)critical temperature, the fixed-scale approach suffers from large lattice artifacts due to small $N_t$, while the fixed-$N_t$ approach can keep $N_t$ finite and reproduces the Stephan-Boltzmann limit with sufficiently large $N_t$.
On the other hand, at small $T$, typically $T$ \lsim\ $T_c$, the fixed-scale approach can keep $a$ small at the price of larger cost due to large $N_t$, while the fixed-$N_t$ approach suffers from lattice artifacts due to small $a$.
See \cite{WHOT09} for more pros and cons.

We test the fixed-scale approach armed with the $T$-integration method in quenched QCD. 
Results are summarized in Fig.\ref{fig:fixedscale}.
Comparing EOS' obtained on various lattices as well as the result from the fixed-$N_t$ approach on large lattices, we find that the fixed-scale approach is reliable and powerful to calculate EOS, in particular at low and intermediate temperatures.
The systematic errors  due to the interpolation in $T$ is well under control in these cases.

\section{$2+1$ flavor QCD with improved Wilson quarks}
\label{sec:Nf21}

   \begin{figure}[b,t]
       \centerline{
    \includegraphics[width=46mm]{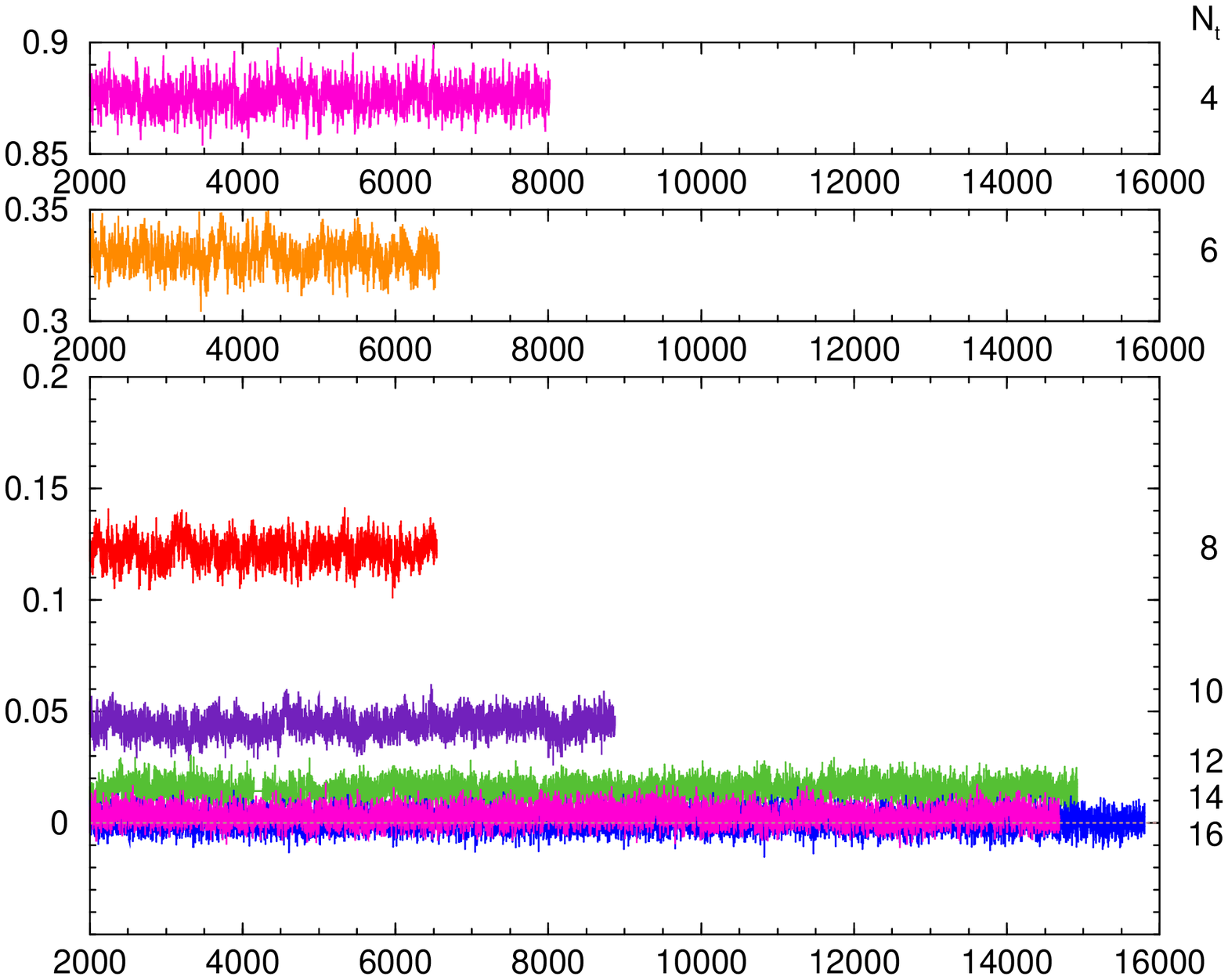}
    \hspace{1mm}
    \includegraphics[width=44mm]{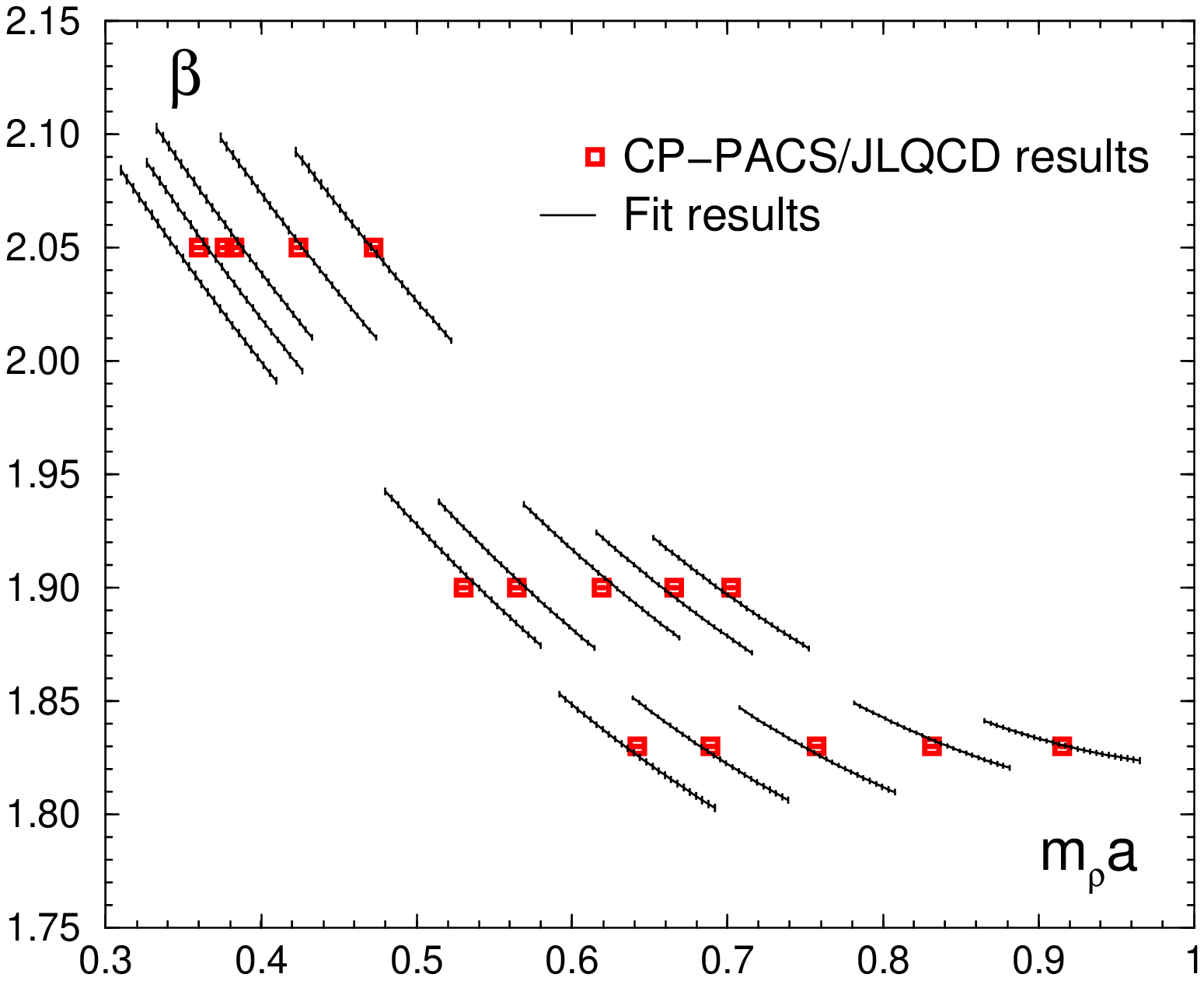}
    \hspace{0.5mm}
    \includegraphics[width=44mm]{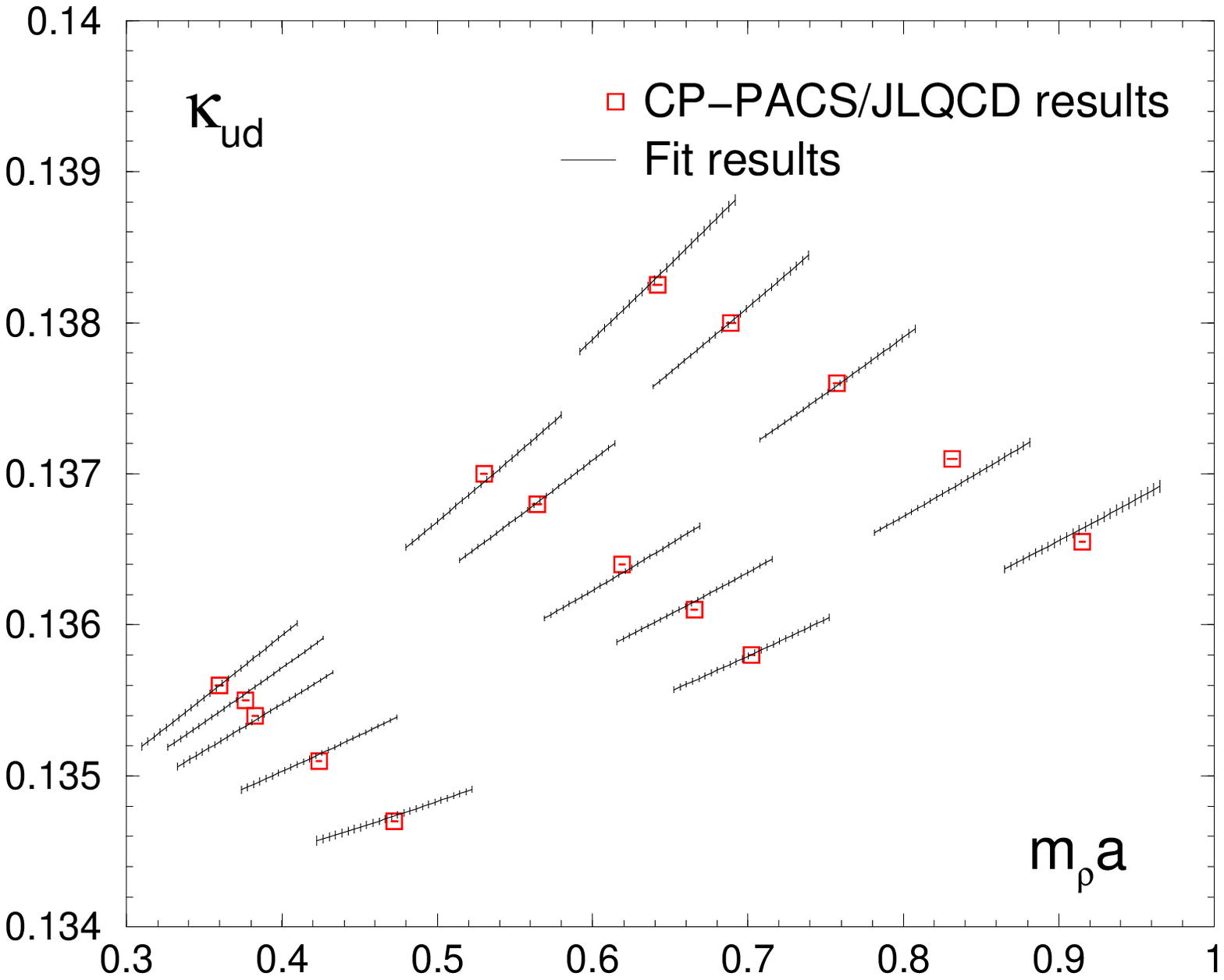}
       }
   \caption{2+1 flavor QCD with  improved Wilson quarks.\cite{WHOT-Lat10}
    {\em Left:} Polyakov loop history.
    {\em Center:} fit result for $\beta$. 
    {\em Right:} the same as the center panel but for $\kappa_{ud}$.}
   \label{fig:Nf21beta}
   \end{figure}

Adopting the fixed-scale approach, we are carrying out a calculation of EOS with $2+1$ flavors of non-perturbatively improved Wilson quarks coupled to RG-improved Iwasaki glue.\cite{WHOT-Lat10} 
As the basic $T=0$  configurations, we use those by the CP-PACS+JLQCD Collaborations \cite{CP-PACS-JLQCD} which are public on the ILDG. 
The spatial lattice volume is about (2 fm)$^3$.
Among their 30 simulation points we have chosen the finest lattice ($a=0.07$ fm, $\beta=2.05$ on a $28^3\times56$ lattice) with the lightest u and d quarks, $m_\pi/m_\rho =0.6337(38)$ and $m_K/m_{K^*} = 0.7377(28)$ for a test study of EOS.
Using the same coupling parameters, we are generating finite temperature configurations on $32^3\times N_t$ lattices with $N_t=4$, 6, $\cdots$ 16 which correspond to $T\sim 170$--700 MeV.
Current status 
is shown in Fig.\ref{fig:Nf21beta}(left). 

To calculate the beta functions, we perform a global fit of the CP-PACS+JLQCD data for $a m_\rho$, $m_\pi/m_\rho$, and $m_{\eta_{ss}}/m_\phi $ available at 30 data points ($3\beta\times 5\kappa_{ud} \times 2\kappa_s$).
We adopt the direct fit method\cite{CPPACS01}: we fit $\beta$, $\kappa_{ud}$ and $\kappa_s$ as polynomial functions of $(a m_\rho, m_\pi/m_\rho, m_{\eta_{ss}}/m_\phi )$ with 10 free parameters each.
Fit results for $\beta$ and $\kappa_{ud}$ are shown in the center and right panels of Fig.\ref{fig:Nf21beta}.
Fit result for $\kappa_s$ is similar.
Although $\chi^2/N_{DF}\sim 2$--5 is marginal, the fits look reasonable.
Taking derivatives along a LCP, we obtain preliminary values $\left( -0.334(4), 0.00289(6), 0.00203(5) \right)$
for the beta functions ${\displaystyle \left( a\frac{d\beta}{da},  a\frac{d\kappa_{ud}}{da},  a\frac{d\kappa_s}{da} \right)}$ at our simulation point, where errors are statistical.

   \begin{figure}[b,t]
       \centerline{
    \includegraphics[width=44mm]{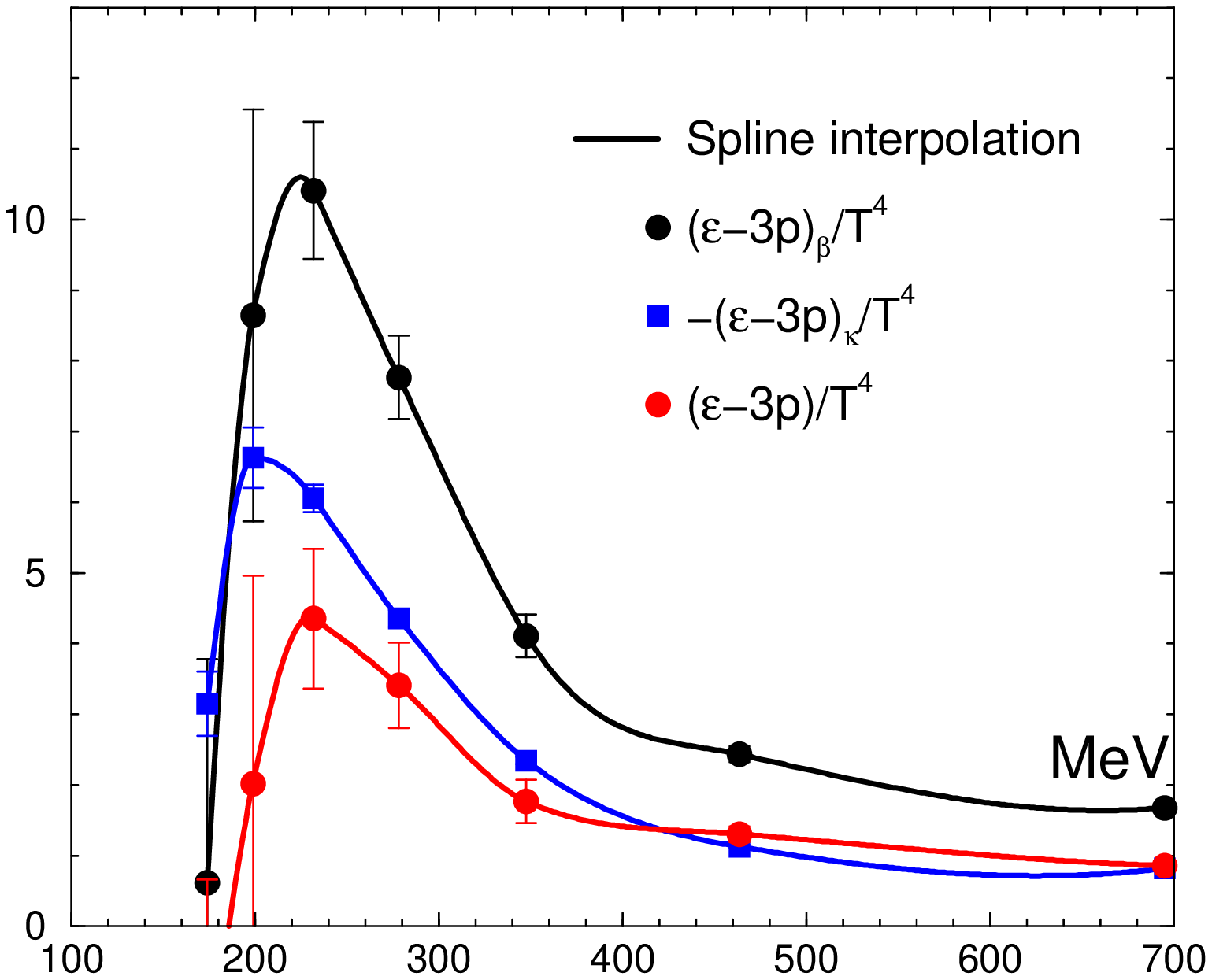}
    \hspace{0.5mm}
    \includegraphics[width=44mm]{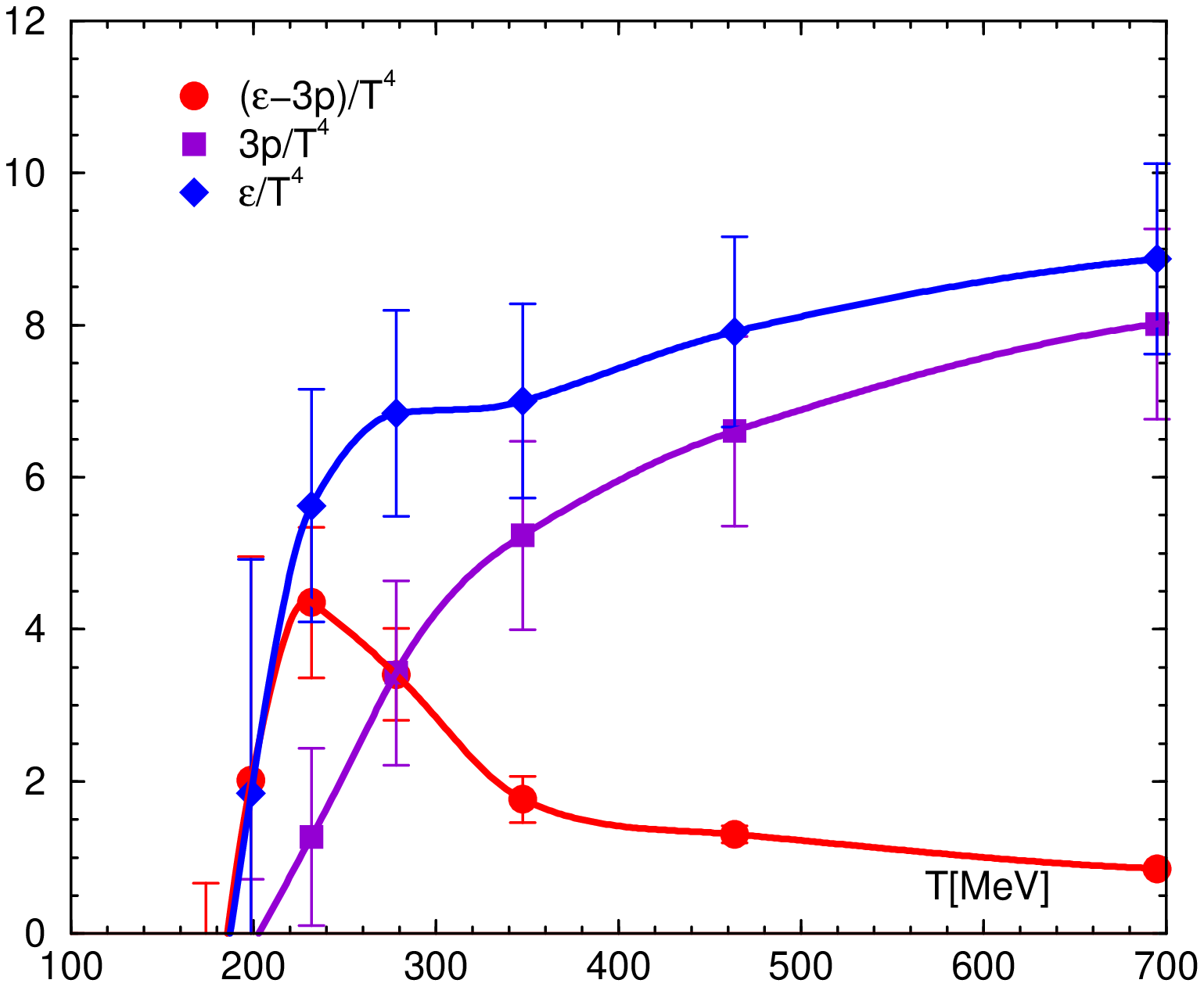}
    \includegraphics[width=53mm]{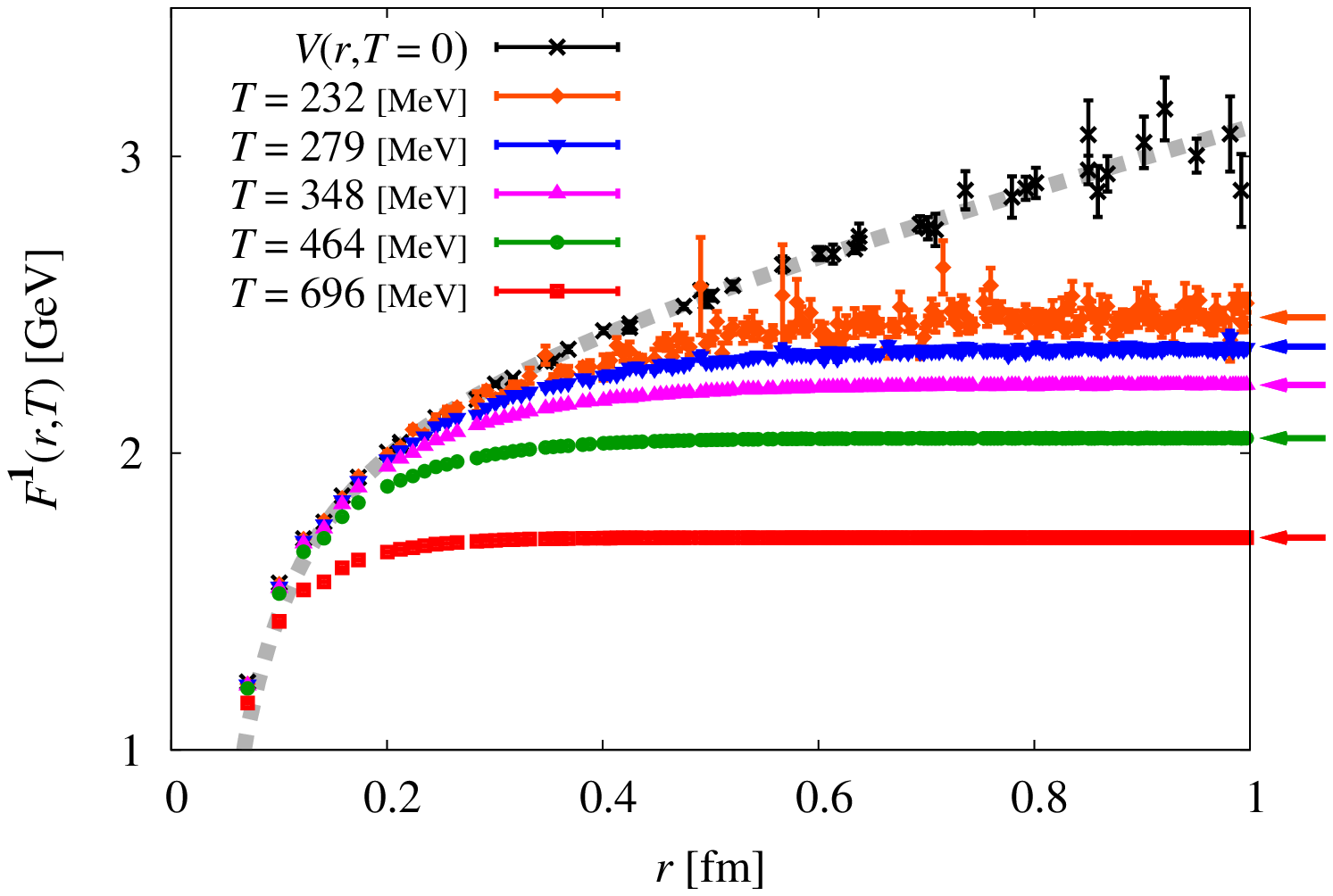}
       }
   \caption{Preliminaly result of EOS for 2+1 flavor QCD with  improved Wilson quarks by the fixed-scale approach. \cite{WHOT-Lat10}
    {\em Left:} trace anomaly. The lattice acale was set by $r_0=0.5$ fm.
    {\em Center:} EOS by the $T$-integration method using the trapezoidal interpolation of the trace anomaly.
    {\em Right:} heavy-quark free-energy in the deconfined phase\cite{MaezawaLat09}.
   The heavy-quark potential $V(r)$ at $T = 0$ was calculated by the CP-PACS+JLQCD Collaborations\cite{CP-PACS-JLQCD} from Wilson-loop expectation values. 
   The arrows on the right side denote twice the single-quark free energy, 
   $2F_Q = - 2T \ln\langle{\rm Tr}\Omega\rangle$}
   \label{fig:Nf21eos}
   \end{figure}

Using these beta functions, we calculate 
$(\epsilon - 3p) = (\epsilon - 3p)_\beta + (\epsilon - 3p)_\kappa$ 
with
\begin{eqnarray}
\frac{(\epsilon - 3p)_\beta}{T^4} &\equiv& \frac{N_t^3}{N_s^3}\, a \frac{d\beta}{da} \left\langle \frac{\partial S}{\partial\beta} \right\rangle ,
\\
\frac{(\epsilon - 3p)_\kappa}{T^4} &\equiv& \frac{N_t^3}{N_s^3}\, \left(
a \frac{d\kappa_{ud}}{da} \left\langle \frac{\partial S}{\partial\kappa_{ud}} \right\rangle 
+ 
a \frac{d\kappa_{s}}{da} \left\langle \frac{\partial S}{\partial\kappa_{s}} \right\rangle 
\right) .
\label{eq:e3p}
\end{eqnarray}
Our results are shown in Fig.\ref{fig:Nf21eos}(left).
We find a relatively small value of about 4 for the peak height of $(\epsilon - 3p)/T^4$.
We note that recent results from highly improved staggered-type quarks (HISQ\cite{HISQ} and stout\cite{stout}) on $N_t=6$--8 lattices give similar small values of 4--5.
Because our lattices with $N_t = 12$--14 around the peak are fine too, we think that our results are consistent with theirs.
Carrying out the $T$-integration (\ref{eq:Tintegral}) using a trapezoidal interpolation of the trace anomaly, we obtain  the pressure $p$ shown in Fig.\ref{fig:Nf21eos}(center).
The energy density $\epsilon$ is calculated from $p$ and $\epsilon -3p$.

Finally, we study the heavy quark free energy in the color singlet channel defined by the Polyakov line correlation function in the Coulomb gauge\cite{MaezawaLat09},
$ 
F^1 (r, T) = - T \ln \langle  {\rm Tr} \Omega^\dag({\bf x}) \Omega({\bf y})   \rangle
$ 
where $r = \left| {\bf x} - {\bf y} \right|$. 
From Fig.\ref{fig:Nf21eos}(right), we find that $F^1(r, T)$ at all temperatures converge to the zero-temperature potential $V(r)$ at short distances, in accordance with the expectation that the short distance physics is insensitive to temperature. 
We stress that, unlike the case of the conventional fixed-$N_t$ approach in which this insensitivity is assumed and used to adjust the constant term of $F^1(r, T)$ at each $T$, we made no such adjustments because the renormalization factors are common to all temperatures in the fixed-scale approach.
We have thus directly confirmed the theoretical expectation.
At large distances, $F^1(r, T)$ departs from $V(r)$ and eventually becomes flat due to Debye screening.
We confirm that $F^1(r, T)$ converges to twice the single-quark free energy quite accurately.

\section{Conclusions and perspectives}
\label{sec:conclusions}

Our final objective is to investigate thermal $2+1$ flavor QCD directly at the physical point.
Corresponding zero-temperature configurations are being generated by the PACS-CS Collaboration applying a reweighting technique to fine-tune to the physical point\cite{PACS-CS}.
With the fixed-scale approach, we can perform finite-temperature simulations directly at the reweighted point.
An extension of the study to finite densities is planned too.
On the other hand, from the tests discussed in the previous section, we found several issues to be solved.
First, a high statistics is required at $T$ \lsim\ $T_c$.
This is due to large cancellations by the zero-temperature subtraction at large $N_t$. 
However, we think that the required computational resources are within the reach. 
Second, 
a finer resolution in $T$ is preferable for EOS at low $T$.
To solve the problem, we are considering to combine results at slightly different $a$ 
taking advantage that our simulations are close to the continuum limit.
Our efforts in other directions were reported by S.\ Ejiri, Y.\ Maezawa and H.\ Saito at the workshop.


This work was presented in the YITP workshop on ``New Frontier in QCD 2010''.
We thank the organizers for their support.
This work is in part supported 
by Grants-in-Aid of the Japanese Ministry
of Education, Culture, Sports, Science and Technology, 
(Nos.~17340066, 18540253, 19549001, 20340047, 21340049)
and by the Grant-in-Aid for Scientific
Research on Innovative Areas (No. 2004: 20105001, 20105003).
This work is in part supported 
by the Large-Scale Numerical Simulation
Projects of CCS, 
Univ.~of Tsukuba,
and by the Large Scale Simulation Program of High Energy Accelerator Research Organization 
(Nos.06-19, 07-18, 08-10, 09-18).

%

\end{document}